# VTXO: the Virtual Telescope for X-ray Observations

John Krizmanic

University of Maryland Baltimore County, Center for Space Science & Technology (CRESST) – NASA/Goddard Space Flight Center
Greenbelt MD 20771 USA, (301) 286-6817
John.F.Krizmanic@nasa.gov

Neerav Shah, Alice Harding, Phil Calhoun, Lloyd Purves, Cassandra Webster

NASA/Goddard Space Flight Center
Greenbelt MD 20771 USA, (301) 286-8174
neerav.shah-1@nasa.gov

Steven Stochaj, Kyle Rankin, Daniel Smith, Hyeongjun Park, Laura Boucheron, Krishna Kota

New Mexico State University
Las Cruces NM 88003 USA, (575) 646-4828
sstochaj@nmsu.edu

Michael Corcoran, Chris Shrader

The Catholic University of America, CRESST-NASA/GSFC
Greenbelt MD 20771 USA, (301) 286-5576
michael.f.corcoran@nasa.gov

Asal Naseri

Space Dynamics Laboratory
Logan UT 84341, (435) 713-3400
asal.naseri@sdl.usu.edu

**ABSTRACT**

The Virtual Telescope for X-ray Observations (VTXO) will use lightweight Phase Frensel Lenses (PFLs) in a virtual X-ray telescope with 1 km focal length and with nearly 50 milli-arcsecond angular resolution. Laboratory characterization of PFLs have demonstrated near diffraction-limited angular resolution in the X-ray band, but they require long focal lengths to achieve this quality of imaging. VTXO is formed by using precision formation flying of two SmallSats: a smaller, 6U OpticsSat that houses the PFLs and navigation beacons while a larger, ESPA-class DetectorSat contains an X-ray camera, a charged-particle radiation monitor, a precision star tracker, and the propulsion for the formation flying. The baseline flight dynamics uses a highly-elliptical supersynchronous geostationary transfer orbit to allow the inertial formation to form and hold around the 90,000 km apogee for 10 hours of the 32.5-hour orbit with nearly a year mission lifetime. The guidance, navigation, and control (GN&C) for the formation flying uses standard CubeSat avionics packages, a precision star tracker, imaging beacons on the OpticsSat, and a radio ranging system that also serves as an inter-satellite communication link. VTXO's fine angular resolution enables measuring the environments nearly an order of magnitude closer to the central engines of bright compact X-ray sources compared to the current state of the art. This X-ray imaging capability allows for the study of the effects of dust scattering nearer to the central objects such as Cyg X-3 and GX 5-1, for the search for jet structure nearer to the compact object in X-ray novae such as Cyg X-1 and GRS 1915+105, and for the search for structure in the termination shock of in the Crab pulsar wind nebula. The In this paper, the VTXO science performance, SmallSat and instrument designs, and mission description is be described. The VTXO development was supported as one of the selected 2018 NASA Astrophysics SmallSat Study (AS3) missions.

## MILLI-ARCSECOND ANGULAR RESOLUTION X-RAY ASTRONOMY

As in other physics disciplines, advances in X-ray astronomy have been driven by the capability to explore Nature on increasingly finer spatial scales. This is exemplified by what was achieved by going from the arcminute resolution view of the X-ray sky provided by the Einstein X-ray telescope [1]} to that provided by the Chandra X-ray Observatory [2] which currently resolves cosmic X-ray sources on arcsecond scale. Einstein Observatory, Chandra Observatory, and even proposed next-generation missions, such as the proposed Lynx X-ray Observatory [3], are limited to angular resolution of 0.5 milli-arcsecond (mas) due to using conventional Wolter type-1 X-ray optics. Breaking this X-ray imaging barrier requires an entirely different X-ray imaging technology, and we have developed X-ray Phase Fresnel Lenses (PFLs) that have been shown to image near the diffraction limit in the X-ray band [4]. Figure 1 shows a SEM of a 3-mm diameter PFL, designed to image at 8 keV with 110-m focal length (left) and the results on imaging characterization (right), demonstrating near diffraction-limited imaging in the X-ray band [5].

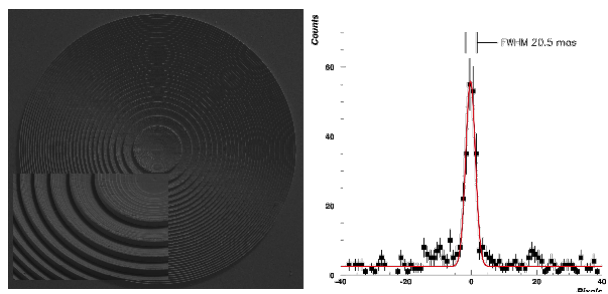

**Figure 1:** Left, an SEM of a 3 mm diameter PFL designed to image at 8 keV, inset shows magnified part of the PFL; Right, the measured point-spread-function of 20.5 mas which is near the diffraction limit value of 16 mas. From Ref. [5]

PFLs designed for VTXO provide the imaging optics in the Virtual Telescope for X-ray Observations (VTXO) SmallSat mission that is designed to achieve nearly 50 mas imaging of astrophysical sources in the X-ray band. The long focal lengths inherent to diffractive PFL optics require VTXO to use two precision formation-flying spacecraft to form the X-ray telescope with 1 km focal length with around 55 mas (FWHM) X-ray imaging resolution, which is nearly 10-times improvement over that provided by Chandra. VTXO's extraordinary angular resolution enables for the first time the exploration of high-energy environments around black holes, neutron stars, and stellar systems at the smallest spatial scales achievable as seen in X-rays.

In principle, PFLs can image at even the micro-arcsecond (μas) scale in the X-ay and gamma-ray energy bands and with meter-size optics, but at the cost of having focal lengths >> 1 km. Thus, VTXO is a also pathfinder mission for a potential mas or (μas) X-ray or gamma-ray missions with square meter photon collecting areas.

## VTXO SCIENCE

VTXO's fine angular resolution enables measuring the environments close to the central engines of compact X-ray sources, allowing for the study of the effects of dust scattering halos nearer to the compact objects in Cygnus X-3, GX 5-1 & Centaurus X-3, the search for dust scattering echoes and jet structures nearer to the central object in X-ray transients such as V404 Cygni, the search for sub 0.1 arcsecond structures in the plerion nebula around the Crab pulsar and in the wind environments ambient to bright X-ray binaries such as Cyg X-1, and imaging of bow shocks in colliding wind binaries, such as Eta Carinae. Sco X-1 will be used as a bright, point source to validate the X-ray imaging performance of VTXO when the SmallSats are in formation. In the development of the VTXO science portfolio, we also identified X-ray observations that would be enabled with an order of magnitude improvement in angular resolution and sensitivity. These include resolving structure in the enigmatic Be star Gamma Cassiopeiae, resolving X-ray sources in regions of high stellar density, and allow for studying the effects of space weather on nearby exoplanets such as Proxima B.

**Table 1: Observation time for 1000 VTXO counts in the energy band 4.5 ± 0.075 keV assuming a 3 cm diameter PFL with 30% collection efficiency.**

| Source | Flux (mCrabs) | Observation Time (hr) |
|---|---|---|
| Sco X-1 | 8000 | 0.2 |
| GX 5-1 | 1260 | 1.5 |
| GRS 1915+105 | 450 | 4.2 |
| Cyg X-3 | 390 | 4.9 |
| Cyg X-1 | 350 | 5.4 |
| Crab Pulsar | 100 | 19 |
| Cen X-3 | 90 | 21 |
| γCas | 13 | 146 |
| Eta Carinae | 4.2 | 452 |

The VTXO science objectives are based on perform 50-mas scale imaging in the X-ray band of the environments around compact astrophysical sources, with the sensitivity defined by the size of the PFL optics. The design of the PFL optics are based on what has been fabricated and characterized at GSFC, specifically PFLs with 3-cm diameter in silicon. The imaging performance and efficiency of PFLs are energy dependent. For X-ray energies above 4 keV, absorptive losses are minimal in silicon and lead to PFL efficiencies of 30% as demonstrated in laboratory tests [5,6}. This motivates the choice of 4.5 keV for one PFL, a low energy with minimal absorption losses to maximize the VTXO count rate assuming power law spectra from the X-ray source. Furthermore, PFLs are inherently chromatic in that their near diffraction-limited imaging is performed over a narrow energy band. Table 1 details the VTXO count rates for the science targets for a PFL imaging at 4.5 ± 0.075 keV assuming 30\% imaging efficiency, For a 1 crab strength source, VTXO acquires 1000 counts in 1.9 hours in the band 4.5 ± 0.075 keV and in 4.9 hours for the band 6.7 ± 0.075 keV, again assuming 30% imaging efficiency. The 6.7 keV energy for the second PFL based on the 6.7-keV emission line from helium-like iron. VTXO will also carry a third PFL-achromat optic to perform extended X-ray bandwidth measurements of the sources.

## PHASE FRESNEL LENS PERFORMANCE

PFLs employ diffraction to focus incident radiation to a primary focal point and can be used as the optics of an X-ray or gamma-ray telescope with near diffraction-limited angular resolution performance [4,7,8]. The focal length is defined by $f = R\, p_{MIN}\} / \lambda$ where R is the PFL radius, $p_{MIN}\}$, is the pitch of the outermost Fresnel zone, and $\lambda$ is the wavelength to be imaged. In practice, the diameter of the PFL is determined by the minimum $p_{MIN}\}$ that can be achieved with the fabrication technique, the energy to be imaged, and the focal length. The maximal height of the Fresnel ridges is material dependent, and the thickness for a $2\pi$ phase change is given by $t_{2\pi} = \lambda / \delta$ where $\delta$ is the index of refraction decrement of the material.

There are three terms that determine the angular resolution for PFLs [7,8]: a the diffraction limit term ($\theta_{Diff} = 1.22\, \lambda / d$), a finite pixel size term $\theta_{Pxl} = \Delta_X / f$, and a chromatic aberration term $\theta_{CA} = 0.2\ (\delta E/E)\ (d/f)$ where d is the diameter of the PFL, $\Delta_X$ is the linear pixel size in the X-ray camera, f is the focal length, E is the energy to be imaged, and $\delta E$ is the energy resolution of the X-ray camera.

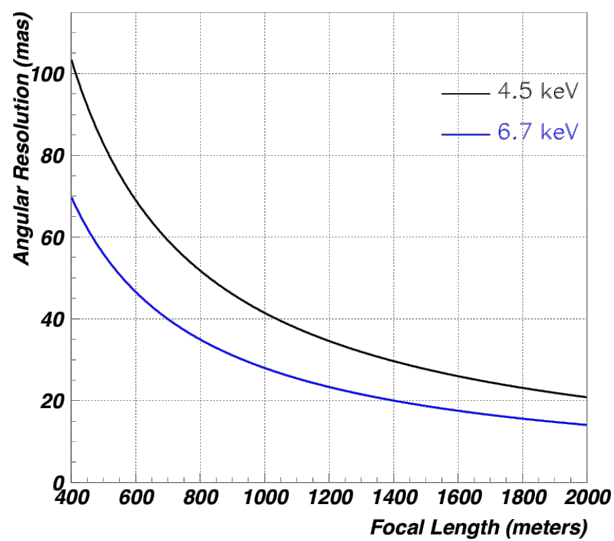

**Figure 2: The angular resolution of VTXO X-ray telescope (PFL optics and H2RG sensor) versus focal length at 4.5 keV and 6.7 keV assuming a 3 cm diameter PFL and 150 eV energy resolution of the X-ray camera.**

For the PFL parameters for VTXO, the chromatic aberration term dominates and the angular resolution for all three terms combined in quadrature is shown in Figure 2 for at 4.5 keV and 6.7 keV as a function of focal length of the telescope. For a focal length of 1 km, the PSFs (FWHM) are 42 mas at 4.5 keV and 28 mas at 6.7 keV. As discussed below, the virtual telescope alignment errors dominate yields and effective angular resolution of 55 mas (FWHM). It should be noted that imaging with PFLs are fairly insensitive to modest tilt angles, alleviating the need to highly control the orientation of the PFL and X-ray camera. Also control of the focal length is relatively modest as the imaging error this introduces is effectively an energy error and just needs to be smaller than the other terms, e.g. a 10 m distance error over 1000 m corresponds to a 0.1% energy error which is much smaller than that in the chromatic aberration term.

## VTXO INSTRUMENTS

**X-ray Camera**: VTXO will use a silicon H2RG HyVisi sensor for the X-ray camera. The sensor will be cooled to 250 K via a single stage thermal electric cooler (TEC). We currently have a silicon H2RG HyVisi in hand and are planning to use this sensor in VTXO. baseline design. The H2RG readout will use an ACADIA ASIC and board developed in-house at GSFC for the WFIRST mission [9]. The VTXO X-ray camera module will contain a baffle and either a beryllium or graphite window to provide optical/UV light shielding as well as to suppress X-rays below ~ 2 keV. The VTXO X-ray camera will be mounted on an optical bench along with the NISTEx-II star tracker in the DetectorSat to minimize any alignment errors due to vibration and thermal effects.

**Trapped Particle Radiation Detector**: While the VTXO orbit was chosen with a 90,000 km altitude apogee that spends more than 50% of the orbit period outside the outer trapped electron radiation belt, we have included a simple trapped particle radiation detector system to validate the radiation environment during science observations. An Amptek Si-SDD sensor, similar to what is used as the X-ray detectors in NICER, is mounted on the DetectorCraft near the opening for the X-ray camera. Amptek A225 and A206 hybrid electronic chips form a charged-particle counter that uses an FPGA to provide the data to the computer. These electronics have been used in the University of Minnesota's High Altitude X-ray Detector Testbed (HAXDT) experiment [10] flown on the High Altitude Student Platform (HASP).

## VTXO PERFORMANCE

The field of view (FoV) of the virtual telescope is set by the linear size of the X-ray camera ($D_{FP}$ = 3.7 cm) for the single Teledyne H2RG HyViSI sensor used for the focal plane, and the focal length (f) of the telescope: FoV} = $D_{FP}$ /f = 8 arcminutes for 1 km focal length. VTXO Flight dynamics studies showed that a 1 km focal length is achievable for the baseline VTXO orbit (90,000 km apogee and 600 km perigee altitudes) with a reasonable amount of propellant for an ESPA-class SmallSat using high-TRL cold gas propulsion leading to a 200+ day mission [11]. The VTXO flight dynamics and GN&C algorithms, denoted as the Formation Flying Control System (FFCS), determined the combined errors on the angular resolution from the formation flying alone This error is dominated by the 41 mas (1-σ) of the NISTEx-II stat tracker that is mounted on an optical bench with the H2rG X-ray camera in the DetectorSat and by the knowledge of the location of the laser beacons on the OpticsCraft. The FFCS instrument alignment errors adding in quadrature to the PFL PSF angular resolution for 1 km focal length yields 55 mas (FWHM) total angular resolution for the VTXO virtual telescope.

### VTXO Mission Conops

Figure 3 illustrates VTXO in formation viewing the Crab nebula. The VTXO is envisioned to be launched to the supersynchronous orbit via a rideshare on a launch similar to the SpaceX FH 2 flight, which achieved the 90,000 km apogee. The VTXO SmallStats will be deployed from an ESPA-ring with the 6U deployed from a standard 6U canister. After initial spacecraft self-checkout, the spacecraft will be put into a loose formation and the perigee will be raised to 600 km after which spacecraft and formation-flying commissioning will occur. The formation separation will be variable around the orbit, with the separation being ~20 m near perigee (to minimize propulsion).

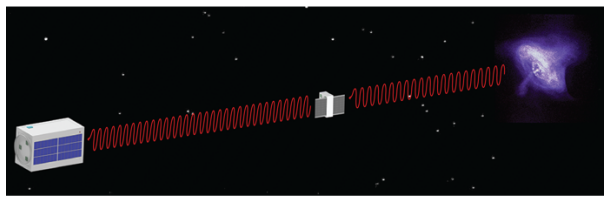

**Figure 3: VTXO in formation viewing the Crab nebula. The DetectorSat is in the foreground while the OpticsSat is in the line of sight to the Crab.**

The SmallSat separation will increase once leaving perigee to achieve the precise science formation 5 hours before apogee and hold this formation for 10 hours. During science mode operation, the precision of the formation is maintained by FFCs control of th DetecotrSat thrusters that keep transverse alignment of the two SmallSats to ± 5 mm. At 5 hours past apogee the formation relaxes with the spacecraft separation decreasing to 20 m when approaching perigee. Ground communications occur at an altitude ~10,000 km before perigee. Figure 4 provides a schematic of the science mode ConOps.

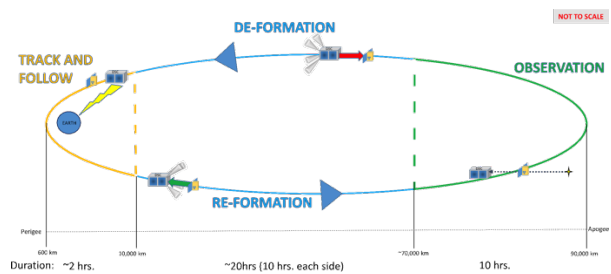

**Figure 4: VTXO concept of operations for one orbit in science mode.**

### VTXO SMALLSAT SPACECRAFT

Referring to Figures 3 and 4, in science precision formation-flying mode the SmallSats are separated by 1 km with the formation controlled by the FFCS. Here we detail the components of the two spacecraft:

**DetecorSat:**

- ESPA-class: 50 × 40 × 80 cm$^3$
- Dry Mass: 72 kg
- Wet Mass: 109 kg
- Power:48 W

- **Instruments:** H2RG X-ray camera and NISTEx-II star tracker on optical bench; charged particle detector.
- Avionics
- VACCO cold gas propulsion: (Δv = 100 m/s)
- S-band radio for ground com and inter-satellite com and radio ranging
- GPS

**OpticsSat:**

- 6U CubeSat
- Dry Mass: 10 kg
- Wet Mass: 12 kg
- Power 24 W
- **Instruments:** 3 PFL optics, laser navigation beacons for formation flying
- Avionics
- VACCO cold gas propulsion: (Δv = 40 m/s)
- S-band radio for ground com and inter-satellite com and radio ranging
- GPS

The telemetry requirements are ~ 5 kbps for the inter-satellite link, needed to exchange navigational data, and < ~200 Mbit/orbit with the maximum determined by the brightest astrophysical target. Radiation dose analysis determined that 5 mm of aluminum shielding is needed to keep the total ionization dose (TID) < 20 kRad/year. Further details for the VTXO instruments and work on formation=flying orbit optimization is found in Ref. [11,12,13,14].

*Acknowledgments*
This work was supported under NASA research announcement, NNH18ZDA001N-AS3 via proposal 18-AS318-0027 at NASA/GSFC, grant 80NSSC19K0123 at University of Maryland, Baltimore County (UMBC) , and grant 80NSSC19K0695 at New Mexico State University (NMSU) and under NASA Cooperative Agreement Notice NNH15ZHA003C at NMSU by grant NM-NNX15AM73A.